\documentclass[dvips]{article}
\usepackage{icrctc07}

\title{Time Evolution of Cosmic Ray MHD Shocks and Their Emissions}
\shorttitle{MHD Cosmic Ray Shocks}
\authors{P. P. Edmon$^{1}$, T. W. Jones$^{1}$, H. Kang$^{2}$.}
\shortauthors{P. P. Edmon et al}
\afiliations{$^1$Department of Astronomy, University of Minnesota, Minneapolis, MN, USA\\ $^2$ Department of Earth Sciences, Pusan National University, Pusan 609-735, Korea}
\email{edmonpp@msi.umn.edu}

\abstract{We present results of time evolution of oblique MHD plane
shocks including diffusive cosmic ray acceleration with  backreaction
on the plasma flows. The simulations include self-consistent effects 
of finite Alfv\'en wave propagation and dissipation.
From the computed cosmic ray particle phase space distributions 
we calculate expected leptonic and hadronic emissions resulting from 
interactions between the cosmic rays, magnetic fields, the thermal 
particle population and relevant astrophysical photon fields.}

\begin{document}
\maketitle

\section{Introduction}
Cosmic ray acceleration in strong shocks is highly efficient
and naturally leads to substantial modification in the shock
structure compared to ordinary gas or MHD shocks described by
Rankine-Hugoniot relations. The modified shock compression can greatly
exceed that of an adiabatic gas shock and the shock structure includes a
foot or precursor where plasma is compressed and heated as it flows
against cosmic rays (CRs) streaming ahead of the relatively thin dissipative shock transition.
Since the cosmic rays CRs are accelerated
by diffusive propagation through the entire shock structure, these shock
modifications also alter the spectrum of the CRs compared to
the test particle spectrum formed in a discontinuous transition.

The presence of magnetic fields is essential to the physics of
diffusive shock acceleration (DSA), because the principal
CR scattering mechanism is gyroresonant interaction with
Alfv\'en waves. That is typically modeled by way of the spatial
diffusion coefficient for the CRs in an otherwise gasdynamic model
of the shock (e.g., \cite{ref2,ref10}). In some calculations the influence of finite streaming
of the Alfv\'en waves with respect to the bulk plasma and
the local dissipation of wave energy (i.e., ``Alfv\'en transport'')
have been included (e.g., \cite{ref5,ref6,ref11}).
On the other hand, despite the facts that typically
one expects an oblique magnetic field with respect to the 
shock normal and that the inclusion of significant
Alfv\'en transport effects imply significant MHD effects,
very few calculations have been published that include a full
MHD treatment of DSA, especially when the CR spectrum is calculated
self-consistently \cite{ref3,ref12}. To explore the importance of
such a self-consistent treatment we include all of the above physics
in the calculations reported here. Since modified CR shocks
are generally evolving structures so long as the CR spectrum
continues to extend to higher energies, our treatment is also
time dependent.

The CRs accelerated in astrophysical shocks will produce
observable electromagnetic emissions through their interactions
with the local plasma and ambient radiation fields. The intensity
and spectra of these emissions will generally depend on the
CR spectrum as well as the structure of the shock. To illustrate the
importance of MHD effects in these emissions we include 
calculations of emissions generated by both leptonic and
hadronic CRs in the modified MHD shocks we present here.

\section{Methods and Model Parameters}
We carried out our simulations in one spatial dimension, $x$,
using our ``Coarse Grained finite Momentum
Volume'' (CGMV) scheme for solving the CR diffusion-convection equation
\cite{ref4}. 
The CGMV scheme evolves the first two momentum
moments of the CR momentum distribution function, $f(t,p,x)$, over finite
momentum bins, $\Delta \ln{p} \sim 1$, assuming a piecewise
powerlaw momentum dependence for $f(t,p,x)$. The powerlaw
slope in each bin is part of the obtained solution. We have demonstrated
that this approach provides accurate solutions to the dynamics
and the evolution of the CR distribution at greatly reduced
computational effort in comparison to finite difference methods. 
The CGMV routines were
incorporated into our well-tested TVD MHD code \cite{ref9} and CR modified
shock solutions were tested against a chain of previously published
simulations.

\begin{figure}
\begin{center}
\vskip-0.3cm
\noindent
\includegraphics [width=0.40\textwidth]{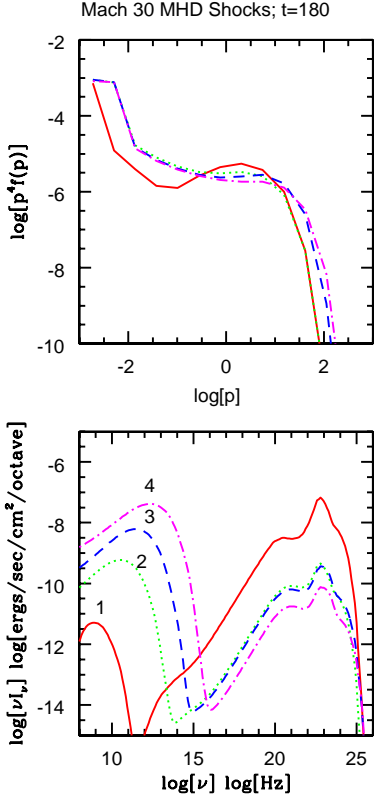}
\end{center}
\caption{
Top: Immediate-postshock CR proton spectra at the gas subshock in
four MHD shocks as outlined in the text and in Table 1. Bottom:
Intensity of EM emissions from the computed shocks
including synchrotron, inverse Compton, bremsstrahlung and
secondary pion decays. Numbers correspond to models in Table 1.
}\label{fig1}
\end{figure}

We assume upstream of the dissipative subshock that
the net scattering turbulence velocity with respect to the plasma
is the Alfv\'en velocity
parallel to the local magnetic field vector in response to resonant
amplification of upstream-facing Alfv\'en waves by streaming CRs.
Downstream of the subshock we assume isotropic Alfv\'enic
turbulence. We assume also that the Alfv\'enic turbulence dissipation
rate matches the local growth rate produced by resonant scattering;
i.e., $- v_A \cdot \nabla P_c = -v_{Ax}\partial P_c/\partial x$ \cite{ref5}. 
The simulations include both a dynamically important CR proton and a 
passive CR electron component. Both components have the same spatial 
diffusion model; the electrons differ in their evolution only through
their energy losses to synchrotron emission and inverse Compton emission
of a combined cosmic microwave background and galactic interstellar
radiation field. Spatial CR
diffusion is isotropic and ``Bohm-like '', resulting in an effective
diffusion coefficient along the shock normal, 
$\kappa = \kappa_0 (B_0/B)p$, with $\kappa_0 = (1/3)(m_pc^3/(eB_0)$. 
We henceforth express all particle momenta in units of $m_p c$.
Subscripts $0$ refer to conditions far upstream. 

Initially the upstream CR population is void. CR protons and electrons
are injected at the subshock 
following the simple presription that a fraction $\epsilon_{inj}$ of
the thermal particle population passing through the shock
is injected into the CR population with momenta 
$p_{inj} = m_p \lambda c_{s,2}$, where $c_{s,2}$ is the postshock 
sound speed, and we set $\lambda = 2$.
In the simulations presented here
the proton injection fraction is  $\epsilon_{inj,p} = 10^{-2}$, 
while the electron value is $\epsilon_{inj,e} = 10^{-4}$.

\begin{table}[t]
\begin{center}
\begin{tabular}{l|ccccc}
\hline
Model & $M_{A,x}$ &       $\theta_0$       & $B_0(\mu G)$ &  $x_{D,0}(cm)$\\
\hline
 1         & 3000     & 0 &   0.46       & 2.3(14)  \\
 2     & 30 & 0     & 46.0     & 2.3(12)    \\
 3     & 30 & 30     & 53.0     & 2.0(12)   \\
 4     & 30 & 75     & 180.0     & 5.7(11)    \\
\hline
\end{tabular}
 \caption{Shock Model Parameters}

\end{center}
\end{table}

These shocks all have initial sonic Mach numbers, 
$M_{s,0} = u_s/c_{S,0} = 30$, with physical shock speed,
$u_s = 0.01c = 3000~{\rm km/s}$.
The large scale magnetic field is placed in the $x-y$ plane.
We include two parallel shocks ($B_y = 0$, $\theta = tan^{-1}{B_{y,0}/B_x} = 0$)
with Alfv\'enic Mach numbers, 
$M_A = u_s/v_{Ax} = 3000,~{\rm and}~30$. 
There are also two oblique MHD shocks
with $M_{Ax} = u_s/v_{Ax}=30$ and $\theta = 30^{\circ}$, and $75^{\circ}$. 
The upstream plasma density, 
$\rho_0 = m_p n_0 = 1.67\times10^{-24}~{\rm g/cm}^3$.
For a given Alfv\'enic Mach number this fixes the upstream magnetic
field strength, as given in Table 1.
Models 2-4 nominally all have the same Alfv\'en wave advection and
dissipation properties.

\section{Results}

The properties of the four simulations at $t = 180$ are 
presented in Figures 1 and 2. By this time each of the shocks
is close to reaching its asymptotic postshock compression and
pressure values, although the shock transition and the CR spectra
would continue to spread self-similarly \cite{ref4}.
The characteristic diffusion length, $x_{D,0} = \kappa_0/u_s$, and diffusion time, $t_{D,0} = \kappa_0/u_s^2$,
along with the upstream mass density, $\rho_0$ provide convenient scaling units.
Setting $\kappa_0 = 1$, we have $u_s = 1$ and an upstream
gas pressure, $P_{g,0} = 1/1500$.
The magnetic field is presented in units such that the magnetic
pressure, $P_B = (1/2) B^2$. The full computational box in each case
spans the spatial domain $[-100,100]$, with the associated
physical $x_{D,0}$ listed in Table 1.

\begin{figure*}
\begin{center}
\vskip-0.5cm
\noindent
\includegraphics [width = 0.28\textwidth,angle=-90]{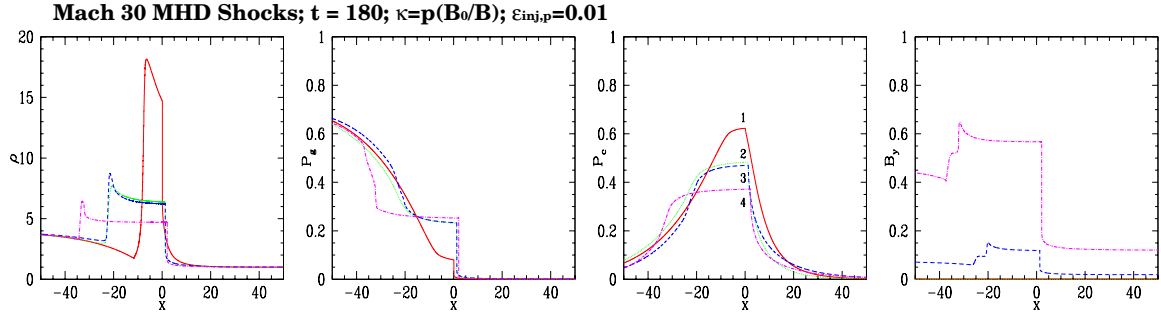}
\end{center}
\caption{Profiles of the four MHD shocks. Left to right:
Gas density, $\rho$, and pressure, $P_g$, CR pressure, $P_c$,  and the 
transverse magnetic field, $B_y$.
Numbers in the $P_c$ plot correspond to models in Table 1.}
\label{fig2}
\end{figure*}

The four shocks are shown at equivalent times in the 
sense that test particle DSA theory would predict the same maximum
CR momentum, $p_{max} \sim (1/8) t \sim 20$ ($E_{max} \sim 20 {\rm GeV}$) (e.g., \cite{ref7}). 
While that prediction
is roughly confirmed in Figure 1, it is also clear that the particle
distributions, the predicted emissions and the shock structures
show obvious differences among the models. To understand the differences
it is easiest to begin with a comparison of the shock structures
as illustrated in Figure 2. Here we see that the total shock compression
is significantly reduced in all the models with $M_{Ax} = 30$ in comparison
to the $M_{Ax} = 3000$ case. That is mostly the result of increased
Alfv\'en wave dissipation in the shock precursor and a reduced net
velocity change sensed by CRs across the shock (e.g., \cite{ref7}). That also
leads to a reduced efficiency in CR acceleration, as pointed out
by a number of previous authors (e.g., \cite{ref8}). There is a further reduction
in shock compression and DSA efficiency when the magnetic field is 
oblique, because the transverse magnetic field component contributes a 
significant pressure gradient that resists compression through the shock. 
The total shock compression in the $\theta = 75^{\circ}$ model 4 is only 1/3
that of the parallel, essentially gasdynamic model 1, and the
postshock $P_c$ is reduced by about a factor of two in the same
comparison.
The particle spectra respond to these trends through a reduction
in the concavity below $p_{max}$, since the reduced compression
reduces the spread in velocity changes sensed by particles as they
scatter through the precursor. We note for these model parameters
and simulation times that the electron and proton CR distributions
do not differ substantially.

Although the differences in particle spectra exhibited in Figure 1
are relatively small, they translate into substantial differences
in the expected electromagnetic spectra. This is illustrated in
the right panel of Figure 1, where we present the intensity of
radiation from synchrotron, inverse Compton, bremsstrahlung and 
secondary pion decay processes found by integrating 
along a line of sight parallel to the shock normal.
The listed order of processes corresponds 
to the order of dominance in features seen in the intensity plot
beginning at low frequencies. The large variation in
synchrotron intensity reflects the two orders of magnitude
range in magnetic field. More interesting, perhaps, is the large
range in the bremsstrahlung and pion decay emissions coming from
the large reductions in shock compression and DSA efficiency
in the MHD shocks.


\section{Conclusions}
MHD CR shocks evolve in noticeably different ways in comparison to
gasdynamic CR shocks. Finite Alfv\'en speeds reduce the efficiency
of diffusive shock acceleration. Magnetic pressure gradients through
the shock transition reduce compression, further reducing acceleration
efficiency. These effects can substantially alter predictions of
the nonthermal emissions associated with the shocks, in particular
reducing nonthermal X-ray and $\gamma$-ray emissions.

\section{Acknowledgements}
This work is supported at
the University of Minnesota by NASA and by the University of Minnesota
Supercomputing Institute and at Pusan National University by KOSEF through
the Astrophysical Research Center for the Structure and Evolution of
the Cosmos (ARCSEC).

\bibliography{icrc0789}
\bibliographystyle{plain}

\end{document}